\documentstyle[aps,psfig]{revtex}
\begin{document}
\twocolumn

\title{Canonical Statistics of Trapped Ideal and Interacting Bose Gases}

\author{Hongwei Xiong$^{1,2}$, Shujuan Liu$^{1}$, Guoxiang Huang$^{3}$, and Zaixin Xu%
$^{3}$}

\address{$^{1}$Department of Applied Physics, Zhejiang University of Technology,
 Hangzhou, 310032, China}
\address{$^{2}$Zhijiang College,
Zhejiang University of Technology, Hangzhou, 310012, China }
\address{$^{3}$Department of Physics, East China Normal University, Shanghai, 200062,%
\\
China}

\date{\today}

\maketitle

\begin{abstract}

The mean ground state occupation number and condensate
fluctuations of interacting and non-interacting Bose gases
confined in a harmonic trap are considered by using a canonical
ensemble approach. To obtain the mean ground state occupation
number and the condensate fluctuations, an analytical description
for the probability distribution function of the condensate is
provided directly starting from the analysis of the partition
function of the system. For the ideal Bose gas, the probability
distribution function is found to be a Gaussian one for the case
of the harmonic trap. For the interacting Bose gas, using a
unified approach the condensate fluctuations are calculated based
on the lowest-order perturbation method and on Bogoliubov theory.
It is found that the condensate fluctuations based on the
lowest-order
perturbation theory follow the law $\left\langle \delta ^{2}N_{{\bf {0}}%
}\right\rangle \sim N$, while the fluctuations based on Bogoliubov
theory behave as $N^{4/3}$.

\end{abstract}


\narrowtext


\section{Introduction}


The experimental achievement of Bose-Einstein condensation (BEC) in dilute
alkali atoms \cite{ALK}, spin-polarized hydrogen \cite{MIT} and recently in
metastable helium \cite{HEL} has enormously stimulated the theoretical
research \cite{RMP,PARKIN,LEG} on the ultracold bosons. Among the several
intriguing questions on the statistical properties of trapped interacting
Bose gases, the problem of condensate fluctuations $\left\langle \delta
^{2}N_{{\bf {0}}}\right\rangle $ of the mean ground state occupation number $%
\left\langle N_{{\bf {0}}}\right\rangle $ is of central importance. Apart
from the intrinsic theoretical interest, it is foreseeable that such
fluctuations will become experimentally testable in the near future \cite
{NEAR}. On the other hand, the calculations of $\left\langle \delta ^{2}N_{%
{\bf 0}}\right\rangle $ are crucial to investigate the phase collapse time
of the condensate \cite{JAV,WRI}.

It is well known that within a grand canonical ensemble the
fluctuations of
the condensate are given by $\left\langle \delta ^{2}N_{{\bf {0}}%
}\right\rangle =N_{{\bf {0}}}\left( N_{{\bf {0}}}+1\right) $, implying that $%
\delta N_{{\bf {0}}}$ becomes of order $N$ when the temperature
approaches zero. To avoid this sort of unphysically large
condensate fluctuations, a canonical (or a microcanonical)
ensemble has to be used to investigate the fluctuations of the
condensate. On the other hand, because in the experiment the trapped
atoms are cooled continuously from the surrounding, the system can
be taken as being in contact with a heat bath but the total number
of particles in the system is conserved. Thus it is necessary to
use the canonical ensemble to investigate the statistical
properties of the trapped weakly interacting Bose gas.

Within the canonical as well as the microcanonical ensembles, the
condensate fluctuations have been studied systematically in the
case of an ideal Bose gas in a box \cite{HAUGE,FUJI,ZIF,BOR,WIL},
and in the presence of a harmonic trap
\cite{WIL,POL,GAJ,GRO,NAVEZ,BAL,HOL1,HOL2}. Recently, the question
of how interatomic interactions affect the condensate fluctuations
has been an object of several theoretical investigations \cite
{GIO,IDZ,ILLU,MEI,KOC,JAK}. Idziaszek {\it et al.} \cite{IDZ}
investigated the condensate fluctuations of interacting Bose gases
using the lowest-order perturbation theory and a two-gas model,
while Giorgini {\it et al.} \cite{GIO} addressed this problem
within a traditional particle-number-nonconserving Bogoliubov
approach. Recently, Kocharovsky {\it et al.} \cite{KOC} supported
and extended the results of the work of Giorgini {\it et al.}
\cite{GIO} using a particle-number-conserving operator formalism.

Although the condensate fluctuations are thoroughly investigated
in Ref.\cite{GIO,IDZ,ILLU,MEI,KOC}, to best our knowledge up to
now an analytical description of the probability distribution
function for the interacting Bose gas directly from the
microscopic statistics of the system has not been given. Note that
as soon as the probability distribution function of the system is
obtained, it is straightforward to get the mean ground state
occupation number and the condensate fluctuations. The purpose of
the present work is an attempt to provide such an analytical
description of the probability distribution function of
interacting and non-interacting Bose gases based on the analysis
of the partition function of the system.

We shall investigate in this paper the condensate fluctuations of
interacting and non-interacting Bose gases confined in a harmonic
trap. The analytical probability distribution function of the
condensate will be given directly from the partition function of
the system using a canonical ensemble approach. For an ideal Bose
gas, we find that the probability distribution of the condensate
is a Gaussian function. In particular, our method can be easily
extended to discuss the probability distribution function for a
weakly interacting Bose gas. A unified way is given to calculate
the condensate fluctuations from  the lowest-order perturbation
theory and from Bogoliubov theory. We find that different methods
of approximation for the interacting Bose gas give quite different
predictions concerning the condensate fluctuations. We show that
the fluctuations based on the lowest-order perturbation theory
follow the law $\left\langle \delta ^{2}N_{{\bf {0}}}\right\rangle
\sim N$, while the fluctuations based on the Bogoliubov theory
behave as $N^{4/3} $.

The paper is organized as follows. Sec. II is devoted to outline
the canonical ensemble, which is developed to discuss the
probability distribution function of Bose gases. In Sec. III we
investigate the condensate fluctuations of the ideal Bose gas
confined in a harmonic trap. In Sec. IV the condensate
fluctuations of the weakly interacting Bose gas are calculated
based on the lowest order perturbation theory. In Sec. V the
condensate fluctuations due to collective excitations are obtained
based on Bogoliubov theory. Finally, Sec. VI contains a
discussion and summary of our results.


\section{Fluctuations and Mean Ground State Occupation Number of the
Condensate in the Canonical Ensemble}


According to the canonical ensemble, the partition function of the
system with $N$ trapped interacting bosons is given by

\begin{equation}
{Z\left[ N\right] =\sum_{\Sigma _{{\bf {n}}}N_{{\bf {n}}}=N}\exp \left[
-\beta \left( \Sigma _{{\bf {n}}}N_{{\bf {n}}}\varepsilon _{{\bf {n}}%
}+E_{int}\right) \right]},  \label{par1}
\end{equation}
\noindent where $N_{{\bf n}}$ and $\varepsilon _{{\bf n}}$ are
occupation number and energy level of the state ${\bf
{n}}=\{n_{x},n_{y},n_{z}\}$, respectively. $\beta =1/k_{B}T$ and
$\{n_{x},n_{y},n_{z}\}$ are non-negative integers. $E_{int}$ is
the interaction energy of the system. For convenience, by
separating out the ground state ${\bf {n}}={\bf 0}$ from the state
${\bf {n}}\neq {\bf 0}$, we have

\begin{equation}
{Z\left[ N\right] =\sum_{N_{{\bf 0}}=0}^{N}\left\{ \exp \left[ -\beta \left(
E_{{\bf 0}}+E_{int}\right) \right] Z_{0}\left( N,N_{{\bf 0}}\right) \right\}}%
,  \label{par2}
\end{equation}
\noindent where $Z_{0}\left( N,N_{{\bf 0}}\right)$ stands for the
partition function of a fictitious system comprising $N-N_{{\bf
0}}$ trapped ideal non-condensed bosons:

\begin{equation}
{Z_{0}\left( N,N_{{\bf 0}}\right) =\sum_{\sum_{{\bf {n}}\neq {\bf 0}}N_{{\bf
{n}}}=N-N_{{\bf 0}}}\exp \left[ -\beta \sum_{{\bf {n}\neq 0}}N_{{\bf {n}}%
}\varepsilon _{{\bf {n}}}\right] .}  \label{II-function-1}
\end{equation}
Assuming $A_{0}\left( N,N_{{\bf 0}}\right) $ is the free energy of
the fictitious system, we have

\begin{equation}
{A_{0}(N,N_{{\bf 0}})=-k_{B}T\ln Z_{0}(N,N_{{\bf 0}}).}  \label{free-energy}
\end{equation}
The calculation of the free energy $A_{0}\left( N,N_{{\bf
{0}}}\right) $ is nontrivial because there is a requirement that
the number of non-condensed
bosons is $N-N_{{\bf {0}}}$ in the summation of the partition function $%
Z_{0}\left( N,N_{{\bf {0}}}\right) $. Using the saddle-point method
developed by Darwin and Fowler \cite{DAR}, it is straightforward to obtain
a useful relation between the free energy $A_{0}\left( N,N_{{\bf {0}}%
}\right) $ and the fugacity $z_{0}$ of the fictitious $N-N_{{\bf 0}}$
non-interacting bosons

\begin{equation}
{-\beta \frac{\partial }{\partial N_{{\bf {0}}}}A_{0}\left( N,N_{{\bf {0}}%
}\right) =\ln z_{0},}  \label{relation1}
\end{equation}
\noindent where the fugacity $z_{0}$ is determined by

\begin{equation}
{N_{{\bf {0}}}=N-\sum_{{\bf {n}}\neq {\bf {0}}}\frac{1}{\exp \left[ {\beta }%
\varepsilon _{{\bf {n}}}\right] z_{0}^{-1}-1}.}  \label{relation2}
\end{equation}
\noindent We have given a simple derivation of Eqs.
(\ref{relation1}) and (\ref {relation2}) in the Appendix.

Using the free energy $A_{0}\left( N,N_{{\bf {0}}}\right) $, the partition
function of the system becomes

\begin{equation}
{Z\left[ N\right] =\sum_{N_{{\bf 0}}=0}^{N}\exp \left[ q\left( N,N_{{\bf 0}%
}\right) \right]},  \label{par3}
\end{equation}
\noindent where

\begin{equation}
{q\left( N,N_{{\bf 0}}\right) =-\beta \left( E_{{\bf 0}}+E_{int}\right)
-\beta A_{0}\left( N,N_{{\bf 0}}\right).}  \label{qqq}
\end{equation}
\noindent It is obvious that $(1/Z\left[ N\right])\exp \left[
q\left( N,N_{{\bf {0}}}\right) \right]$ represents the probability
finding $N_{{\bf 0}}$ atoms in the condensate.

To obtain the probability distribution function of the system, let
us
first investigate the largest term in the sum of the partition function $%
Z\left[ N\right]$. Assume the number of the condensed atoms is $N_{{\bf 0}%
}^{p}$ in the largest term of the partition function. The largest
term $\exp \left[ q\left( N,N_{{\bf {0}}}^{p}\right) \right] $ is
determined by
requiring that $\frac{\partial }{\partial N_{{\bf {0}}}}q\left( N,N_{{\bf {%
0}}}\right) |_{N_{{\bf {0}}}=N_{{\bf {0}}}^{p}}=0$, {\it i.e.},

\begin{equation}
{-\beta \frac{\partial }{\partial N_{{\bf 0}}^{p}}\left( E_{{\bf 0}%
}+E_{int}\right) -\beta \frac{\partial }{\partial N_{{\bf 0}}^{p}}A_{0}(N,N_{%
{\bf 0}}^{p})=0}.  \label{main1}
\end{equation}
\noindent Using Eq. (\ref{relation1}) we obtain

\begin{equation}
{\ln z_{0}^{p}=\beta \frac{\partial }{\partial N_{{\bf 0}}^{p}}\left( E_{%
{\bf 0}}+E_{int}\right).}  \label{z0p}
\end{equation}
\noindent In addition, from Eq. (\ref{relation2}), the most probable value $%
N_{{\bf 0}}^{p}$ is determined by

\begin{equation}
{N_{{\bf {0}}}^{p}=N-\sum_{{\bf {n}}\neq {\bf {0}}}\frac{1}{\exp \left[
\beta \varepsilon _{{\bf {n}}}\right] \left( z_{0}^{p}\right) ^{-1}-1}.}
\label{non1}
\end{equation}
\noindent In the case of an ideal Bose gas, from Eq. (\ref{z0p})
one
obtains $\ln z_{0}^{p}=\beta \varepsilon _{{\bf {0}}}$. Thus $N_{{\bf {0}}%
}^{p}$ is the same as the mean ground state occupation number
obtained by using a grand canonical ensemble approach. For
sufficiently large $N$, the sum $\sum_{N_{%
{\bf {0}}}=0}^{N}$ in (\ref{par3}) may be replaced by the largest
term, since the error omitted in doing so is statistically
negligible. In this situation, Eq. (\ref{non1}) shows the
equivalence between the canonical ensemble and the grand canonical
ensemble for large $N$.

The other terms in the partition function (\ref{par3}) will
contribute to the fluctuations of the
condensate, and lead to the deviation of $\left\langle N_{{\bf {0}}%
}\right\rangle $ from the most probable value $N_{{\bf 0}}^{p}$. If $N_{%
{\bf 0}}\neq N_{{\bf 0}}^{p}$, we have $\frac{\partial }{\partial N_{{\bf 0}}}%
q\left( N,N_{{\bf 0}}\right)\neq 0$. Assuming

\begin{equation}
{\frac{\partial }{\partial N_{{\bf 0}}}q\left( N,N_{{\bf 0}}\right) = \alpha
\left( N,N_{{\bf 0}}\right),}  \label{q-alpha}
\end{equation}
\noindent from Eqs. (\ref{relation1}) and (\ref{qqq}), we obtain

\begin{equation}
{\ln z_{0}=\beta \frac{\partial }{\partial N_{{\bf {0}}}}\left( E_{{\bf {0}}%
}+E_{int}\right) +\alpha \left( N,N_{{\bf {0}}}\right).}  \label{z00}
\end{equation}
\noindent By Eqs. (\ref{relation2}) and (\ref{z00}), we have

$${
N_{{\bf 0}}=N-
}$$

\begin{equation}
{\sum_{{\bf n\neq 0}}\frac{1}{\exp \left[ \beta \varepsilon _{%
{\bf n}}\right] \exp \lbrack -\beta {\frac{\partial }{\partial N_{{\bf {0}}}}%
\left( E_{{\bf {0}}}+E_{int}\right) -\alpha \left( N,N_{{\bf {0}}}\right) }%
\rbrack -1}.}  \label{main3}
\end{equation}
Combining Eqs. (\ref{non1}) and (\ref{main3}), we get the
following equation for determining  $\alpha \left( N,N_{{\bf
{0}}}\right) $

\[
{N_{{\bf {0}}}\vspace{1pt}-N_{{\bf {0}}}^{p}=\sum_{{\bf n}\neq {\bf 0}}\frac{%
1}{\exp \left[ \beta \varepsilon _{{\bf n}}\right] \exp \lbrack -\beta {%
\frac{\partial }{\partial N_{{\bf {0}}}^{p}}\left( E_{{\bf {0}}%
}+E_{int}\right) }\rbrack -1}}
\]

\begin{equation}
{-\sum_{{\bf n}\neq {\bf 0}}\frac{1}{\exp \left[ \beta \varepsilon _{{\bf n}%
}\right] \exp \lbrack -\beta {\frac{\partial }{\partial N_{{\bf {0}}}}\left(
E_{{\bf {0}}}+E_{int}\right) -\alpha \left( N,N_{{\bf {0}}}\right) }\rbrack
-1}.}  \label{alpha}
\end{equation}
\noindent Once we know $E_{{\bf {0}}}$ and $E_{int}$ of the
system, it is
straightforward to obtain $\alpha \left( N,N_{{\bf {0}}}\right) $ from Eq. (%
\ref{alpha}). Using $\alpha \left( N,N_{{\bf {0}}}\right) $, one can obtain
the probability distribution function of the system.

From Eq. (\ref{q-alpha}), we obtain the following result for $q\left( N,N_{%
{\bf {0}}}\right) $
\begin{equation}
{q\left( N,N_{{\bf {0}}}\right) =\int_{N_{{\bf {0}}}^{p}}^{N_{{\bf {0}}%
}}\alpha \left( N,N_{{\bf {0}}}\right) dN_{{\bf {0}}}+q\left( N,N_{{\bf {0}}%
}^{p}\right).}  \label{qalpha2}
\end{equation}
\noindent Thus the partition function of the system becomes

\begin{equation}
{Z\left[ N\right] =\sum_{N_{{\bf {0}}}=0}^{N}\left\{ \exp \left[ q\left(
N,N_{{\bf {0}}}^{p}\right) \right] G\left( N,N_{{\bf {0}}}\right) \right\},}
\label{par-alpha}
\end{equation}
\noindent where

\begin{equation}
{G\left( N,N_{{\bf {0}}}\right) =\exp \left[ \int_{N_{{\bf {0}}}^{p}}^{N_{%
{\bf {0}}}}\alpha \left( N,N_{{\bf {0}}}\right) dN_{{\bf {0}}}\right] .}
\label{ideal-dis}
\end{equation}
\noindent Assuming $P\left( N_{{\bf {0}}}|N\right) $ is the
probability to
find $N_{{\bf {0}}}$ atoms in the condensate, $G\left( N,N_{{\bf {0}}%
}\right) $ represents the ratio $\frac{P\left( N_{{\bf {0}}}|N\right) }{%
P\left( N_{{\bf {0}}}^{p}|N\right) }$, {\it i.e.}, the relative
probability to find $N_{{\bf 0}}$ atoms in the condensate. From
Eq. (\ref{ideal-dis}), the normalized probability distribution
function is given by

\vspace{1pt}
\begin{equation}
{G}_{n}{\left( N,N_{{\bf {0}}}\right) =A\exp \left[ \int_{N_{{\bf {0}}%
}^{p}}^{N_{{\bf {0}}}}\alpha \left( N,N_{{\bf {0}}}\right) dN_{{\bf {0}}%
}\right] ,}  \label{norm-di}
\end{equation}
where $A$ is a normalization constant and is given by the condition $A\int
G(N,N_{{\bf {0}}})dN_{{\bf {0}}}=1$.

As soon as  we know $G\left( N,N_{{\bf {0}}}\right) $, the
statistical properties of the system can be clearly described.
From Eqs. (\ref{par-alpha}) and (\ref
{ideal-dis}) one obtains the mean ground state occupation number $%
\left\langle N_{{\bf {0}}}\right\rangle $ and fluctuations
$\left\langle \delta ^{2}N_{{\bf {0}}}\right\rangle $ in the
canonical ensemble:

\begin{equation}
{\left\langle N_{{\bf {0}}}\right\rangle =\frac{\sum_{N_{{\bf {0}}}=0}^{N}N_{%
{\bf {0}}}\exp \left[ q\left( N,N_{{\bf {0}}}\right) \right] }{\sum_{N_{{\bf
{0}}}=0}^{N}\exp \left[ q\left( N,N_{{\bf {0}}}\right) \right] }=\frac{%
\sum_{N_{{\bf {0}}}=0}^{N}N_{{\bf {0}}}G\left( N,N_{{\bf {0}}}\right) }{%
\sum_{N_{{\bf {0}}}=0}^{N}G\left( N,N_{{\bf {0}}}\right) }}
\label{mean-ideal}
\end{equation}

$${
\left\langle \delta^{2} N_{{\bf {0}}}\right\rangle =\left\langle N_{{\bf {0}%
}}^{2}\right\rangle -\left\langle N_{{\bf {0}}}\right\rangle ^{2}=
}$$

\begin{equation}
{\frac{%
\sum_{N_{{\bf {0}}}=0}^{N}N_{{\bf {0}}}^{2} G\left( N,N_{{\bf {0}}}\right) }{%
\sum_{N_{{\bf {0}}}=0}^{N} G\left( N,N_{{\bf {0}}}\right) }-\left[ \frac{%
\sum_{N_{{\bf {0}}}=0}^{N}N_{{\bf {0}}} G\left( N,N_{{\bf {0}}}\right) }{%
\sum_{N_{{\bf {0}}}=0}^{N} G\left( N,N_{{\bf {0}}}\right) }\right] ^{2}.}
\label{fluc-ideal}
\end{equation}
\noindent Starting from Eqs. (\ref{mean-ideal}) and
(\ref{fluc-ideal}), one can calculate the mean ground state
occupation number and fluctuations for ideal and interacting Bose
gases.


\section{Ideal Bose Gases}


We now study the condensate fluctuations of the system with $N$
non-interacting bosons trapped in an external potential. The
potential is a harmonic one with the form

\begin{equation}
V_{ext}\left( {\bf {r}}\right) =\frac{m}{2}\left( \omega
_{x}^{2}x^{2}+\omega _{y}^{2}y^{2}+\omega _{z}^{2}z^{2}\right) {,}
\label{potential}
\end{equation}
where $m$ is the mass of atoms, $\omega _{x}$, $\omega _{y}$, and $%
\omega _{z}$ are frequencies of the trap along three
coordinate-axis directions. The single-particle energy level has
the form

\begin{equation}
{\varepsilon _{{\bf {n}}}=\left( n_{x}+\frac{1}{2}\right) \hbar \omega
_{x}+\left( n_{y}+\frac{1}{2}\right) \hbar \omega _{y}+\left( n_{z}+\frac{1}{%
2}\right) \hbar \omega _{z}.}  \label{energy-trap}
\end{equation}
From Eq. (\ref{non1}) one can get easily the most probable value
$N_{{\bf {0}}}^{p}$, which reads

\begin{equation}
{N_{{\bf {0}}}^{p}=N-N\left( \frac{T}{T_{c}^{0}}\right) ^{3}-\frac{3%
\overline{\omega }\zeta \left( 2\right) }{2\omega _{ho}\left[ \zeta \left(
3\right) \right] ^{2/3}}\left( \frac{T}{T_{c}^{0}}\right) ^{2}N^{2/3},}
\label{ideal-most}
\end{equation}

\noindent where $T_{c}^{0}=\frac{\hbar \omega _{ho}}{k_{B}}\left( \frac{N}{%
\zeta \left( 3\right) }\right) ^{1/3}$ is the critical temperature
of the ideal Bose gas in the thermodynamic limit. $\overline{\omega
}=\left( \omega _{x}+\omega _{y}+\omega _{z}\right) /3$ and
$\omega _{ho}=\left( \omega _{x}\omega _{y}\omega _{z}\right)
^{1/3}$ are arithmetic and geometric averages of the oscillator
frequencies, respectively. When obtaining (\ref{ideal-most}) we
have used the following expression of the density of states
\cite{FIN}

\begin{equation}
{\rho \left( E\right) =\frac{1}{2}\frac{E^{2}}{\left( \hbar \omega
_{ho}\right) ^{3}}+\frac{3\overline{\omega }E}{2\omega _{ho}\left( \hbar
\omega _{ho}\right) ^{2}}.}  \label{density-state}
\end{equation}

\noindent On the basis of the same density
of states a detailed study of the critical temperature and the
ground state occupation number was given recently in \cite{MULKEN}.

In a thermodynamic equilibrium the
deviation from the most probable value
$N_{{\bf {0}}}^{p}$ is small, therefore we can use the approximation $%
\exp \left[ -\alpha \left( N,N_{{\bf {0}}}\right) \right] \approx
1-\alpha \left( N,N_{{\bf {0}}}\right) $. From Eq. (\ref{alpha})
and the single-particle energy level (\ref
{energy-trap}) we find the result for $\alpha \left( N,N_{{\bf {0}}%
}\right) $

\begin{equation}
{\alpha \left( N,N_{0}\right) =-\frac{\zeta \left( 3\right) \left(
N_{0}-N_{0}^{p}\right) }{\zeta \left( 2\right) N\left( T/T_{c}^{0}\right)
^{3}}.}  \label{alpha-har}
\end{equation}
When obtaining $\alpha \left( N,N_{{\bf {0}}}\right) $ we have
used the expansion $g_{3}\left( 1+\delta \right) \approx \zeta
\left( 3\right) +\zeta \left( 2\right) \delta $ \cite{ROB}, where
$g_{3}\left( z\right) $ belongs to the class of functions
$g_{\alpha }\left( z\right) =\sum_{n=1}^{\infty }z^{n}/n^{\alpha
}$ and $\zeta \left( n\right) $ is Riemann $\zeta $ function. From
 (\ref{ideal-dis}) and (\ref{alpha-har}) we obtain the normalized
probability distribution function of the harmonically trapped
ideal Bose gas

\begin{equation}
{G_{ideal}\left( N,N_{0}\right) =A}_{ideal}{\exp \left[ -\frac{\zeta \left(
3\right) \left( N_{0}-N_{0}^{p}\right) ^{2}}{2\zeta \left( 2\right) N\left(
T/T_{c}^{0}\right) ^{3}}\right] ,}  \label{dis-har}
\end{equation}
where\noindent\ $A_{ideal}$ is a normalization constant. It is
interesting to note  that the expression  (\ref{dis-har}) is a
Gaussian distribution function. From the formulas (\ref
{mean-ideal}), (\ref{fluc-ideal}), (\ref{ideal-most}), and
(\ref{dis-har}) we can obtain  $\left\langle N_{{\bf
{0}}}\right\rangle $ and $\left\langle \delta ^{2}N_{{\bf
{0}}}\right\rangle $ for the ideal Bose gas. 

In Fig. 1(a) and Fig. 1(b) we plot $\left\langle N_{{\bf {0}}}\right\rangle /N$
as a function of temperature for $N=200$ and $N=10^{3}$ ideal bosons
confined in an isotropic harmonic trap. The dashed line displays $%
\left\langle N_{{\bf {0}}}\right\rangle /N$ in the thermodynamic
limit, while the solid line displays $\left\langle N_{{\bf
{0}}}\right\rangle /N$ within the grand canonical ensemble (or
$N_{{\bf {0}}}^{p}$ within the canonical ensemble). The dotted
line displays $\left\langle N_{{\bf {0}}}\right\rangle
/N$ within the canonical ensemble. When $N>10^{3}$, $\left\langle N_{{\bf {0}}%
}\right\rangle /N$ from the canonical ensemble agrees well with
that from the
grand canonical ensemble. Obviously, in the case of $N\rightarrow \infty $, $%
\left\langle N_{{\bf {0}}}\right\rangle /N$ obtained from the
canonical ensemble coincides with that from the grand canonical
ensemble.

From the formulas (\ref{mean-ideal}) and (\ref{fluc-ideal}) and
the results (\ref{ideal-most}) and (\ref{dis-har}) we can obtain
the condensate fluctuations of the ideal Bose gas. In
Fig. 2 we plot the numerical result of $\delta N_{{\bf {0}}}=\sqrt{%
\left\langle \delta ^{2}N_{{\bf {0}}}\right\rangle }$ (solid line) for $%
N=10^{3}$ ideal bosons confined in an isotropic harmonic potential. The
dashed line displays the result of Holthaus {\it et al.} \cite{HOL2}, where
the saddle-point method is developed to avoid the failure of the standard
saddle-point approximation below the onset of BEC.

\begin{figure}[tb]
\psfig{figure=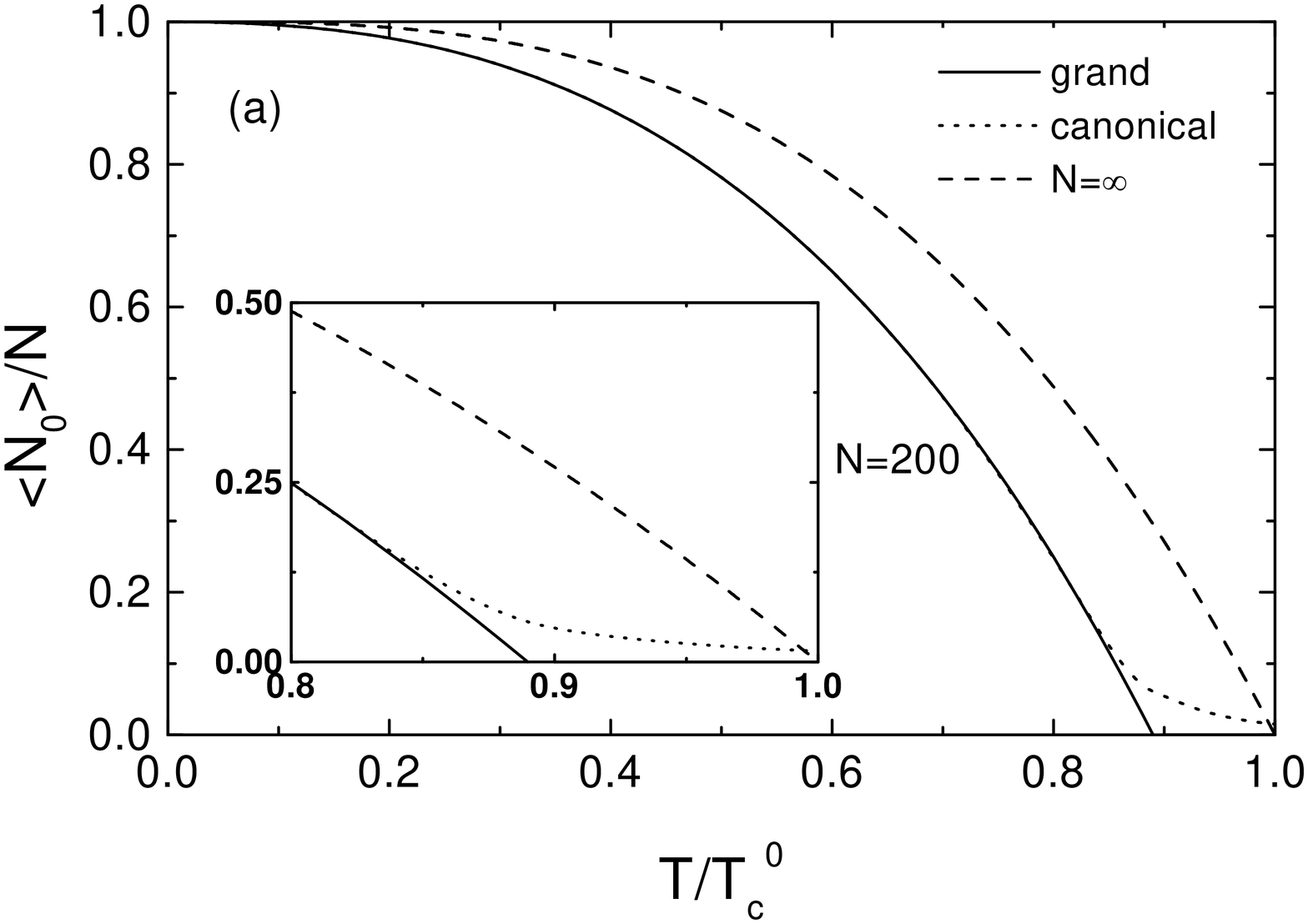,width=\columnwidth,angle=270}
\end{figure}

\begin{figure}[tb]
\psfig{figure=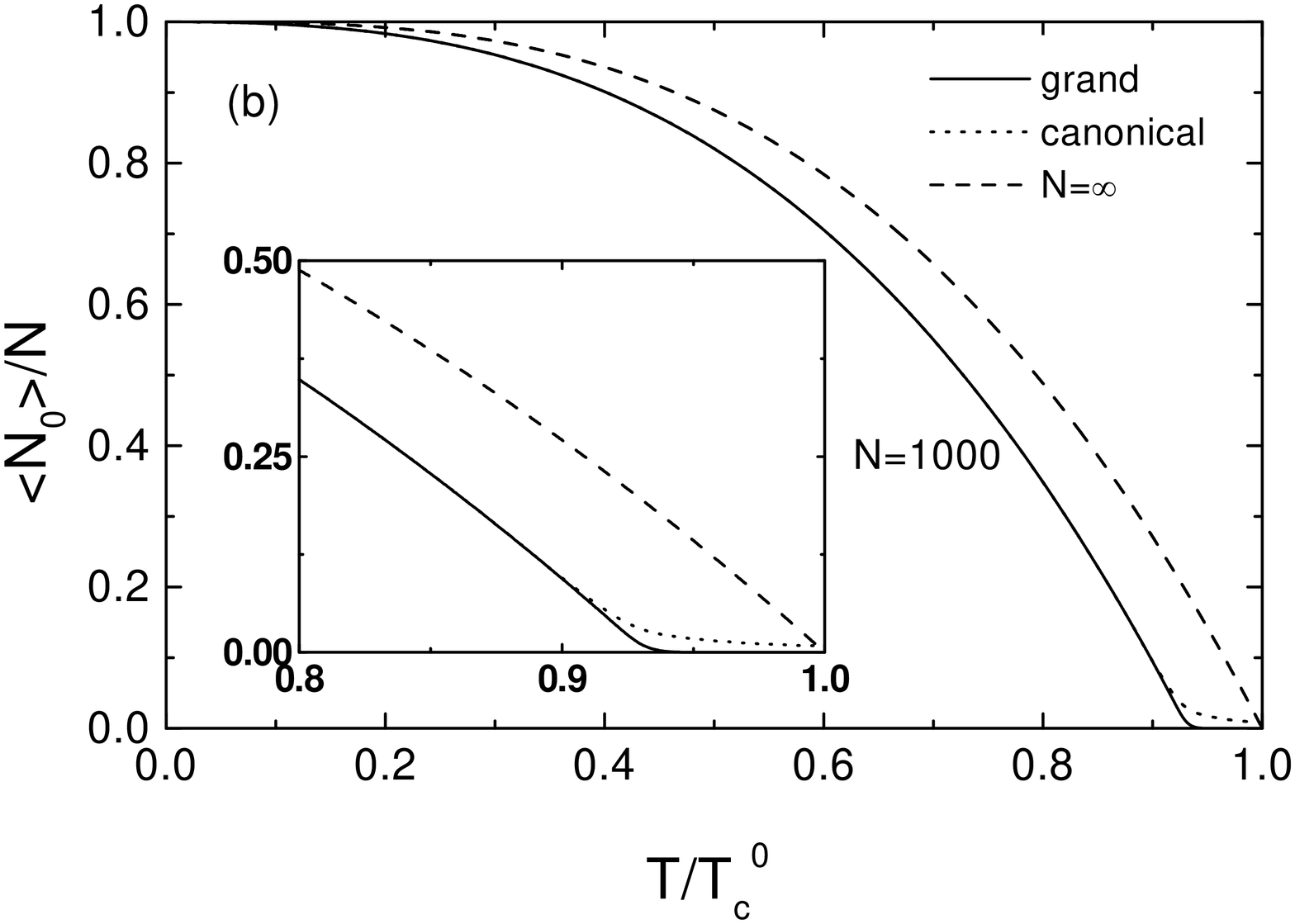,width=\columnwidth,angle=270}
\caption{Relative mean ground state occupation number $\left\langle N_{\bf{0}}\right\rangle /N$
vs $T/T_{c}^{0}$ for $N=200$, $10^{3}$ non-interacting bosons confined in an
isotropic harmonic trap. The dashed line shows $\left\langle N_{\bf{0}}\right\rangle /N$
in the thermodynamic limit.
The solid line shows $\left\langle N_{\bf{0}}\right\rangle /N$ within the grand canonical ensemble,
while the dotted line displays $\left\langle N_{\bf{0}}\right\rangle /N$
within the canonical ensemble.
When $N\rightarrow \infty $, the mean ground state occupation number of the canonical ensemble
coincides with that of the grand canonical ensemble.}
\end{figure}

In Fig. 2 the dotted line displays the result given in
Refs.\cite{POL,GIO}. Our results
coincide with those of Refs.\cite{POL,GIO} when $T/T_{c}^{0}$ is smaller than $%
T_{m}/T_{c}^{0}$, which corresponds to the maximum fluctuations
$\left\langle
\delta ^{2}N_{{\bf 0}}\right\rangle _{\max }$. In fact, when $%
T/T_{c}^{0}<T_{m}/T_{c}^{0}$, from  (\ref{mean-ideal}), (\ref{fluc-ideal}%
) and (\ref{dis-har}), we obtain the analytical result for the
condensate fluctuations:

\begin{equation}
{\left\langle \delta^{2} N_{{\bf 0}}\right\rangle =\frac{\pi
^{2}}{6\zeta \left( 3\right) }N\left( \frac{T}{T_{c}^{0}}\right)
^{3},} \label{analytical}
\end{equation}
which recovers the result given in Refs.\cite{POL,GIO}. This shows
the validity of the probability distribution function (\ref
{dis-har}) for studying the statistical properties of the system.
At the critical temperature, however, our results give

\begin{equation}
{\left\langle \delta ^{2}N_{{\bf 0}}\right\rangle
|_{T=T_{c}}=\left( 1-\frac{2}{\pi} \right) \frac{\pi ^{2}N}{6\zeta
\left( 3\right) },}  \label{critical-fluc}
\end{equation}
\noindent which is much smaller than the result of Ref.\cite{GIO}.
This difference is apprehensible because the analysis of Giorgini
{\it et al.}
\cite{GIO} holds in the canonical ensemble except near and above $T_{c}^{0}$%
, while our result holds also for the temperature near $T_{c}^0$.
Near the critical temperature, our result (solid line) agrees with
that of Holthaus
{\it et al.} \cite{HOL2}. The results given by  (\ref{analytical}) and (%
\ref{critical-fluc}) show a normal behavior of the condensate
fluctuations for the harmonically trapped ideal Bose gas.

\begin{figure}[tb]
\psfig{figure=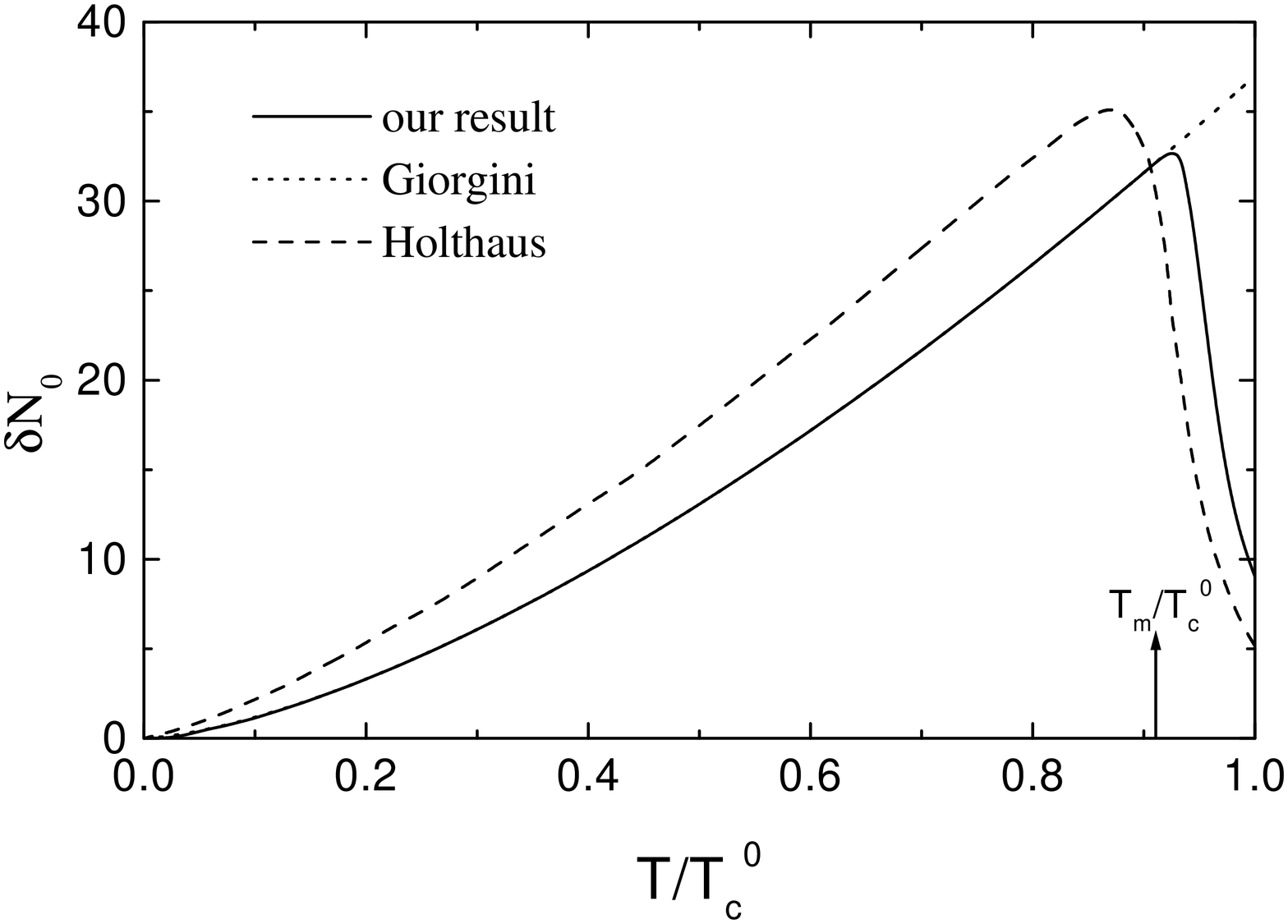,width=\columnwidth,angle=270}
\caption{Root-mean-square fluctuations $\delta N_{\bf{0}}$ for $N=10^{3}$ non-interacting
bosons confined in an isotropic harmonic traps. The solid line displays $\delta N_{\bf{0}}$
obtained from the probability distribution function Eq. (\ref{dis-har}), while the dashed line shows the result
of Holthaus {\it et al.} [21]. The dotted line displays the result of Giorgini {\it et al.} 
[22] ( Eq. (\ref{analytical}) ).
The arrow marks the temperature corresponding to the maximum condensate fluctuations.
Below $T_{m}$, the solid line coincides with the result of Giorgini {\it et al.} [22].
Near $T_{c}$, our result agrees with that of Holthaus {\it et al.} [21].}
\end{figure}

The fluctuations of the condensate can also be evaluated at $T=0$. In the
case of $T\rightarrow 0$, from (\ref{dis-har}) we get $G\left( N,N_{%
{\bf {0}}}\right) =A_{ideal}$ if $N_{{\bf {0}}}=N$, while $G\left( N,N_{{\bf {0}}%
}\right) \rightarrow 0$ when $N_{{\bf {0}}}\neq N$. Therefore, we obtain $%
\left\langle N_{{\bf {0}}}\right\rangle \rightarrow N$ and $\left\langle
\delta ^{2}N_{{\bf {0}}}\right\rangle \rightarrow 0$ when $T\rightarrow 0$.

Note that our results are reliable although the disputable
saddle-point method is used to investigate the fluctuations of the
condensate. It is well known that the applicability of the
saddle-point approximation for the condensed Bose gas has been the
subject of a long debate \cite{ZIF,DEB}. Recently, the analysis
given in Ref.\cite{GRO} showed that the fluctuations are
overestimated, and do not appear to vanish properly with
temperature using the conventional saddle-point method. Our
discussions on the condensate fluctuations are reasonable because
of two reasons: (i) As proved given in Ref.\cite {HOL2}, the most
probable value Eq. (\ref{non1}) for the non-interacting Bose gas
is still correct, even when carefully dealing with the failure of
the standard saddle-point method below the critical temperature.
(ii) In the
usual statistical method $\left\langle N_{{\bf {0}}}\right\rangle $ and $%
\left\langle \delta ^{2}N_{{\bf {0}}}\right\rangle $ are obtained
through the first and second partial derivatives of partition
function, respectively. When the saddle-point approximation is
used to calculate the partition function of the system, the error
will be overestimated in the second partial derivative of the
partition function. Thus one  can not obtain correct condensate
fluctuations using usual method. However, in our approach here
what we used is the reliable result given by Eqs. (\ref{non1}) and
(\ref{main3}). The probability distribution function of the ground
state occupation number can be obtained directly from Eqs.
(\ref{non1}) and (\ref {main3}), without resorting to the second
partial derivative of the
partition function. $\left\langle N_{{\bf {0}}}\right\rangle $ and $%
\left\langle \delta ^{2}N_{{\bf {0}}}\right\rangle $ are obtained from the
probability distribution function in our approach. The correct description of $%
\delta N_{{\bf {0}}}$ near zero temperature and critical
temperature also shows the validity of our method. Thus our
approach has provided in some sense a simple method recovering the
applicability of the saddle-point method through the calculations
of the probability distribution function of the system.


\section{Interacting Bose gases Based on the Lowest Order Perturbation Theory
}

Below the critical temperature, Bose-Einstein condensation results
in a sharp enhancement of the density in the central region of the
trap. This makes the interacting effect between atoms be much more
important than above $T_{c}$. The correction to the condensate
fraction and critical temperature due to the interatomic
interaction has been discussed within grand canonical ensemble
\cite{GIO3,NAR3,BER,LIU} and canonical
ensemble\cite{XIONG,MULKEN}. In this section we investigate the
role of interaction on the condensate fluctuations of a weakly
interacting Bose gas.

Using the lowest-order perturbation theory the interaction energy
of the system takes the form

\begin{equation}
{E_{int}=2g\int n_{0}\left( {\bf {r}}\right) n_{T}\left( {\bf {r}}\right)
d^{3}{\bf {r}}+ g\int n_{T}^{2}\left( {\bf {r}}\right) d^{3}{\bf {r},}}
\label{inter-energy}
\end{equation}
\noindent where $g=4\pi \hbar ^{2}a/m$ is the coupling constant
fixed by the
s-wave scattering length $a$. $n_{0}\left( {\bf {r}}\right) $ and $%
n_{T}\left( {\bf {r}}\right) $ are the density distributions of
the condensate and normal gas, respectively.

Below the critical temperature, by Thomas-Fermi approximation the
density distribution of the condensate reads

\begin{equation}
{n_{0}\left( {\bf {r}}\right) =\frac{\mu -V_{ext}\left( {\bf {r}}\right) }{g}%
,}  \label{density-condensate}
\end{equation}
where $\mu $ is the chemical potential of the system. The
temperature dependence of the chemical potential is then fixed by
the number of atoms in the condensate

\begin{equation}
{\mu (N_{{\bf {0}}},T)=\frac{\hbar \omega _{ho}}{2}\left( \frac{15N_{{\bf {0}%
}}a}{a_{ho}}\right) ^{2/5},}  \label{chemical}
\end{equation}
where $a_{ho}=\left( \hbar /m\omega _{ho}\right) ^{1/2}$ is the
harmonic oscillator length. Moreover, since $\mu =\partial E_{{\bf {0}}%
}/\partial N_{{\bf {0}}}$, the energy per particle in the condensate turns
out to be

\begin{equation}
{\varepsilon _{{\bf {0}}}^{TF}=E_{{\bf {0}}}/N_{{\bf {0}}}=\frac{5}{7}\mu
(N_{{\bf {0}}},T).}  \label{TF-energy}
\end{equation}

As a first-order approximation, omitting the interaction between
condensed and non-condensed atoms, the partition function of the
system is given by

\begin{equation}
{Z_{int}\left[ N\right] =\sum_{N_{{\bf {0}}}=0}^{N} \left\{ \exp \left[
-\beta N_{{\bf {0}}}\varepsilon _{{\bf {0}}}^{TF}\right] Z_{0}\left( N,N_{%
{\bf {0}}}\right) \right\}.}  \label{par-hf}
\end{equation}
From Eq. (\ref{non1}) the most probable value reads

\begin{equation}
{N_{{\bf {0}}}^{p}=N-\sum_{{\bf {n}}\neq {\bf {0}}}\frac{1}{\exp \left[
\beta \left( \varepsilon _{{\bf {n}}}-\mu (N_{{\bf 0}}^{p},T)\right) \right]
-1}.}  \label{most-hf-1}
\end{equation}
\noindent Using the density of states, {\it i.e.},
(\ref{density-state}), one obtains the result for the most
probable value $N_{{\bf {0}}}^{p}$

$${
N_{{\bf {0}}}^{p}=N\left( 1-t^{3}\right) -
}$$

\begin{equation}
{\frac{\zeta \left( 2\right) }{%
\zeta \left( 3\right) } \frac{\mu ( N_{{\bf {0}}}^{p},T) t^{3}N}{k_{B}T}-
\frac{3\overline{\omega }\zeta \left( 2\right) }{2\omega _{ho} \left[\zeta
\left( 3\right)\right] ^{2/3}}t^{2}N^{2/3},}  \label{most-hf}
\end{equation}
\noindent where $t=T/T_{c}^{0}$ is the reduced temperature.
Introducing the scaling parameter $\eta$ \cite{RMP}

\begin{equation}
{\eta =\frac{\mu (T=0)}{k_{B}T_{c}^{0}}=1.57\left( \frac{N^{1/6}a}{a_{ho}}%
\right) ^{2/5},}  \label{eta}
\end{equation}
\noindent  (\ref{most-hf}) becomes

$${
N_{{\bf {0}}}^{p}=N\left( 1-t^{3}\right) -
}$$

\begin{equation}
{\frac{\zeta \left( 2\right) }{%
\zeta \left( 3\right) } \eta Nt^{2}\left( \frac{N_{{\bf {0}}}^{p}}{N}\right)
^{2/5}- \frac{3\overline{\omega }\zeta \left( 2\right) } {2\omega
_{ho}\left[\zeta \left( 3\right) \right]^{2/3}}t^{2}N^{2/3}.}
\label{most-hf-tra}
\end{equation}

\noindent Note that the corrections due to the interatomic
interaction and finite number of particle of the system can be
obtained simultaneously when (\ref{most-hf-tra}) is used to calculate $\left\langle N_{\bf{0}}\right\rangle $ 
and $\left\langle \delta ^{2}N_{\bf{0}}\right\rangle $ of the system.
The second term on the right hand side of
(\ref{most-hf-tra}) accounts for the correction of the interaction
effect. The correction due to the interatomic interaction
coincides with the lowest-order thermal depletion obtained in the
grand canonical ensemble approach \cite{RMP}.

For other $N_{{\bf {0}}}$, assuming $\frac{\partial }{\partial N_{{\bf {0}}}}%
q\left( N,N_{{\bf {0}}}\right) =\alpha \left( N,N_{{\bf
{0}}}\right) $, we get

\begin{equation}
{N_{{\bf {0}}}=N-\sum_{{\bf {n}}\neq {\bf {0}}}\frac{1}{\exp \left[ \beta
\left( \varepsilon _{{\bf {n}}}-\mu (N_{{\bf 0}},T)\right) \right] \exp
\left[ -\alpha \left( N,N_{{\bf {0}}}\right) \right] -1}.}
\label{alpha-hf-1}
\end{equation}
\noindent Combining Eqs. (\ref{most-hf-1}) and (\ref{alpha-hf-1}),
one obtains the result for $\alpha \left( N,N_{{\bf {0}}}\right)
$:

\begin{equation}
{\alpha \left( N,N_{{\bf {0}}}\right) =-\frac{\zeta \left( 3\right) \left(
N_{{\bf {0}}}-N_{{\bf {0}}}^{p}\right) }{\zeta \left( 2\right) Nt^{3}}+\frac{%
\mu ( N_{{\bf {0}}}^{p},T) -\mu ( N_{{\bf {0}}},T) }{k_{B}T}.}
\label{alpha-hf}
\end{equation}
\noindent The probability distribution function of the interacting
Bose gas is then

$${
G_{int}
\left( N,N_{{\bf {0}}}\right)
=A_{int}{\exp \left[ \int_{N_{{\bf {0}%
}}^{p}}^{N_{{\bf {0}}}}\alpha \left( N,N_{{\bf {0}}}\right) dN_{{\bf {0}}%
}\right] =}
}$$

\begin{equation}
\frac{{{A}_{int}}}{A_{ideal}}{G_{ideal}\left( N,N_{{\bf {0}}%
}\right) R_{int}\left( N,N_{{\bf {0}}}\right) ,}  \label{dis-hf}
\end{equation}
\noindent where $A_{int}$ is a normalization constant. $G_{ideal}\left( N,N_{%
{\bf {0}}}\right) $ is the probability distribution function given by the formula
(\ref{dis-har}) for the ideal harmonically trapped Bose gas. The correction $%
R_{int}\left( N,N_{{\bf {0}}}\right) $ originating from the
interatomic interaction takes the form

$${
R_{int}\left( N,N_{{\bf {0}}}\right) =
 \exp \left\{ 
\frac{\hbar \omega _{ho}
}{2k_{B}T}\left( \frac{15a}{a_{ho}}\right) ^{2/5}\times
\right.
}$$

\begin{equation}
{\left. \left[ \left( N_{{\bf {0}}
}^{p}\right) ^{2/5}\left( N_{{\bf {0}}}-N_{{\bf {0}}}^{p}\right)- \frac{5}{7}
\left( N_{{\bf {0}}}^{7/5}-\left( N_{{\bf {0}}}^{p}\right) ^{7/5}\right)
\right] \right\}.}  \label{r-hf}
\end{equation}
\noindent Note that $G_{int}\left( N,N_{{\bf {0}}}\right) $ is not
a Gaussian distribution function because of the existence of the
non-Gaussian factor $R_{int}\left( N,N_{{\bf {0}}}\right) $.

\begin{figure}[tb]
\psfig{figure=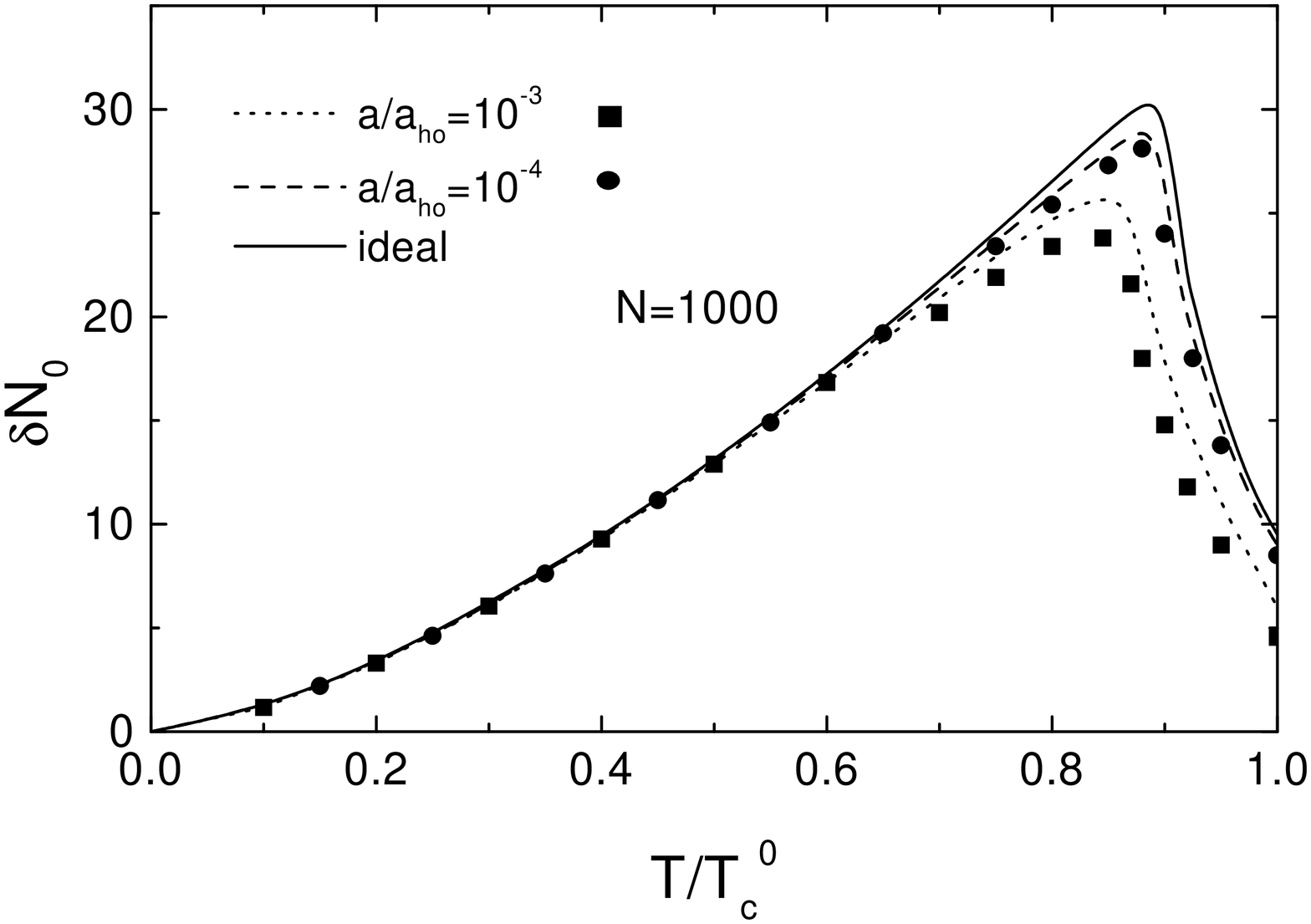,width=\columnwidth,angle=270}
\caption{Root-mean-square fluctuations $\delta N_{\bf{0}}$ for $N=10^{3}$ non-interacting
bosons confined in an isotropic harmonic traps. The solid line displays $\delta N_{\bf{0}}$
obtained from the probability distribution function Eq. (\ref{dis-har}), while the dashed line shows the result
of Holthaus {\it et al.} [21]. The dotted line displays the result of Giorgini {\it et al.} 
[22] ( Eq. (\ref{analytical}) ).
The arrow marks the temperature corresponding to the maximum condensate fluctuations.
Below $T_{m}$, the solid line coincides with the result of Giorgini {\it et al.} [22].
Near $T_{c}$, our result agrees with that of Holthaus {\it et al.} [21].}
\end{figure}

From (\ref{fluc-ideal}) and (\ref{dis-hf}) we can obtain the
numerical result of $\left\langle \delta ^{2}N_{{\bf
{0}}}\right\rangle _{int}$. In Fig. 3 we have provided the
numerical result of $\delta N_{{\bf {0}}}$ for $N=1000$
interacting bosons confined in an isotropic harmonic trap with $%
a/a_{ho}=10^{-4}$ and $a/a_{ho}=10^{-3}$, respectively. The
crossover from the interacting to the non-interacting Bose gases
is clearly shown. From Fig. 3 we find that the repulsive
interaction between atoms results in a decrease of the condensate
fluctuations. For an attractive interaction, we anticipate that
the corrections between atoms result in an increase for the
condensate fluctuations.

The interaction between condensed and non-condensed atoms gives
high-order correction to the thermodynamic properties of the
system. Near
the critical temperature, {\it i.e.}, when $N_{{\bf 0}}a/a_{ho}<<1$ \cite{RMP}%
, we have $n_{0}\left( {\bf {r}}\right) =N_{{\bf 0}}\left( \frac{m\omega _{ho}^{2}}{%
\pi \hbar }\right) ^{3/2}e^{-m\left( \omega _{x}x^{2}+\omega
_{y}y^{2}+\omega _{z}z^{2}\right) /\hbar }$. In addition, we can
adopt the semiclassical approximation for the normal gas
\cite{RMP}, {\it i.e.},
$n_{T}\left( {\bf {%
r}}\right) =\lambda _{T}^{-3}g_{3/2}\left( e^{-\beta V_{ext}\left( {\bf {r}}%
\right) }\right) $ with $\lambda _{T}=\left[ 2\pi \hbar
^{2}/\left( mk_{B}T\right) \right] ^{1/2}$ being the thermal
wavelength. From Eqs. (\ref {non1}) and (\ref{inter-energy}), it
is straightforward to obtain the most probable value $N_{{\bf
0}}^{p}$ near the critical temperature:

\begin{equation}
{\frac{N_{{\bf 0}}^{p}}{N} =\frac{1-t^{3} -\frac{\zeta \left( 2\right) }{%
\zeta \left( 3\right) }\left[ 2-\frac{S}{\zeta \left( 3/2\right) }\right]
\theta t^{7/2} -\frac{3\zeta \left( 2\right) }{2 \left [\zeta \left(
3\right) \right ]^{2/3}} \frac{\overline{\omega }}{\omega _{ho}}t^{2}N^{-1/3}%
} {1+\frac{\zeta \left( 2\right) \theta N^{1/2}t^{2}} {\left
[\zeta \left( 3\right) \right ]^{1/2}\zeta \left( 3/2\right) }}. }
\label{near}
\end{equation}
where $S=\sum_{i,j=1}^{\infty }{1}/{\zeta \left( 3\right) \left[
ij\left( i+j\right) \right] ^{3/2}}$. When obtaining (\ref{near})
we have introduced a scaling parameter $\theta ={%
gn_{T}\left( {\bf {r}}={\bf 0},T_{c}^{0}\right) }/{k_{B}T_{c}^{0}}=2.02\frac{%
a}{a_{ho}}N^{1/6}$. $\theta$ can also be written in the form of
$\theta=0.65\eta^{5/2}$. By setting $N_{{\bf 0}}^{p}=0$, from
(\ref{near}) we obtain the shift of the critical temperature:

\begin{equation}
\frac{\delta T_{c}^{0}}{T_{c}^{0}}= -1.65\frac{a}{a_{ho}}N^{1/6}-\frac{\zeta
\left( 2\right) } {2 \left [\zeta \left( 3\right) \right ]^{2/3}}\frac{%
\overline{\omega }}{\omega _{ho}}N^{-1/3}.  \label{shift}
\end{equation}
The first term on the right hand side of (\ref{shift}) is the
shift due to the interatomic interaction. It agrees with the
results based on the local density approximation \cite{GIO3}
obtained by using  the grand-canonical ensemble approach. The
second term in  (\ref{shift}) gives exactly the usual results due
to effects of the finite number of particles \cite{RMP}. Thus 
in our approach, within the canonical ensemble the corrections due
to the effects of the finite particle number and the interatomic
interactions can be obtained simultaneously.

Below the critical temperature, the most probable value is given by

$${
\frac{N_{{\bf 0}}^{p}}{N} =1-t^{3} - \frac{3\overline{
\omega }\zeta \left( 2\right) }{2\omega _{ho} \left [\zeta \left( 3\right)
\right ]^{2/3}}t^{2}N^{-1/3} 
}$$

\begin{equation}
{-  \frac{\zeta \left( 2\right) }{\zeta
\left( 3\right) }t^{3}\left[ \frac{\eta \xi ^{2/5}}{t}+1.49\frac{\eta t^{2}}{
\xi ^{2/5}}F\left( w\right) +0.14\eta ^{5/2}t^{1/2}\right] 
,}  \label{below1}
\end{equation}
\noindent where $w=\left( \eta \xi ^{2/5}/t\right) ^{1/2}$.
$F\left( w\right) $ is defined by

\begin{equation}
{F\left( w\right) =0.53\left( 1-0.5e^{-0.23w^{3}}-0.5e^{-1.51w^{3}}\right) }.
\label{below2}
\end{equation}
Omitting the high-order terms of the parameter $\eta $, the
expression  (\ref{below1}) gives exactly the lowest-order
correction of (\ref{most-hf-tra}).

\begin{figure}[tb]
\psfig{figure=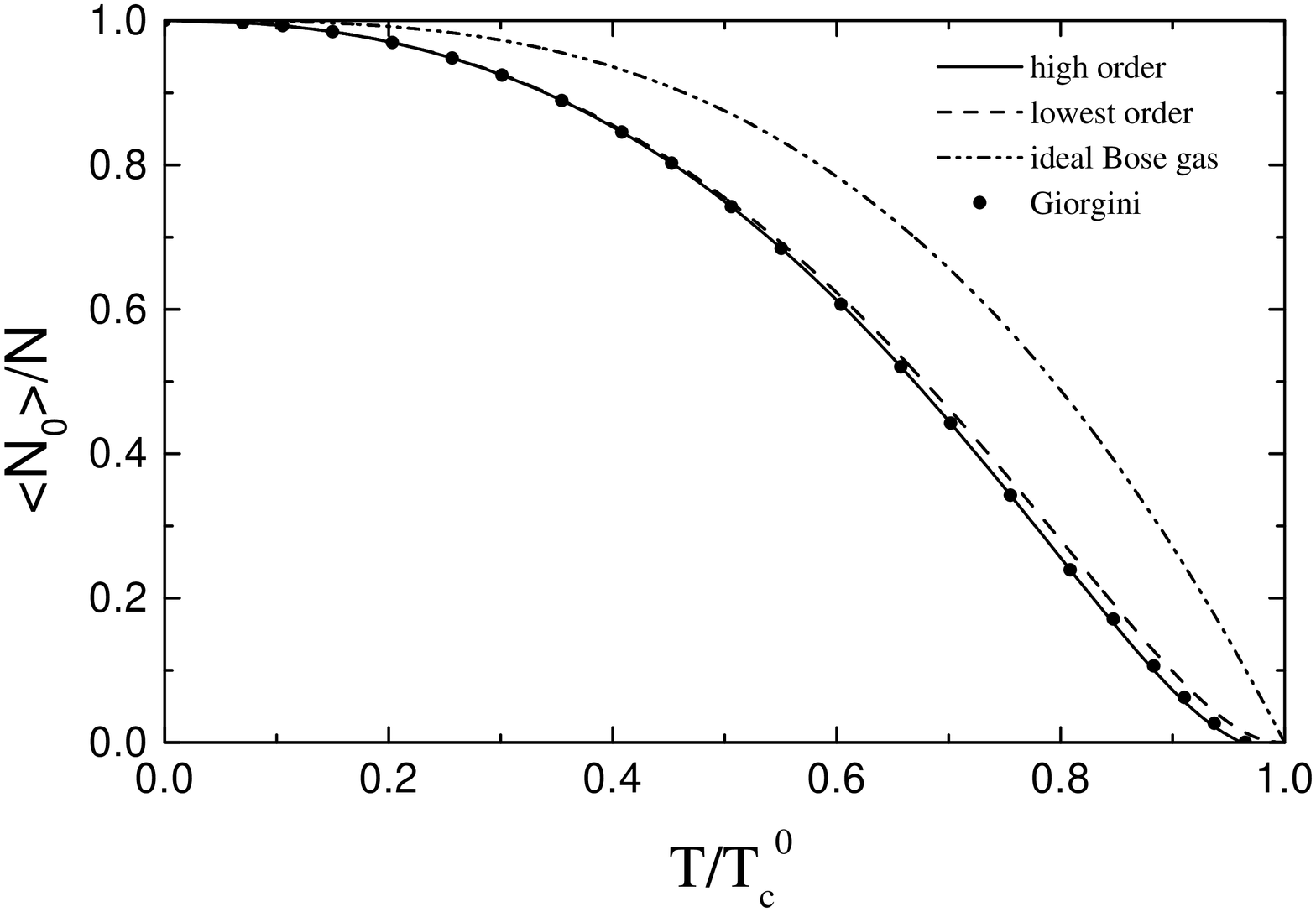,width=\columnwidth,angle=270}
\caption{Displayed is $\left\langle N_{\bf{0}}\right\rangle /N$ of the trapped
interacting Bose gases with the experimental parameters of Ensher {\it et al.} [38].
The dashed-dotted line shows $\left\langle N_{\bf{0}}\right\rangle /N$ of the ideal
Bose gas, while the circles show the result of Giorgini {\it et al.} [33].
The numerical results of the lowest order and high order 
$\left\langle N_{\bf{0}}\right\rangle /N$ within the canonical ensemble
are displayed with dashed and solid lines,
respectively.}
\end{figure}

From  (\ref{inter-energy}) we can obtain the probability
distribution function of the condensate when the interaction
between condensed and non-condensed atoms is considered. Combining
with the most probable value,
one obtains $\left\langle N_{{\bf {0}}}\right\rangle $ and $\delta N_{{\bf {0%
}}}$ of the interacting Bose gases. In Fig. 4 the experimental
parameter by
Ensher {\it et al.} \cite{ENS} is used to plot $\left\langle N_{{\bf {0}}}\right\rangle /N$
within the canonical ensemble. Our results (solid line)
agree well with the conclusion of Ref.\cite{GIO3}
(circles) where semiclassical approximation is used in the frame
of the grand-canonical ensemble. We can also obtain the numerical
results for $\delta N_{{\bf {0}}}$ in the presence of the
interaction between condensed and non-condensed atoms. The
numerical result of $\delta N_{{\bf {0}}}$ is displayed in Fig. 3
with $a/a_{ho}=10^{-4}$ (circles) and $a/a_{ho}=10^{-3}$ (squares),
respectively. Our calculations show that the repulsive
interaction between the condensed and non-condensed atoms lowers
the condensate fluctuations further.


\section{Interacting Bose gases based on Bogoliubov theory}


Condensate fluctuations due to collective excitations have been
recently investigated by Giorgini {\it et al.} \cite{GIO} within
the traditional particle-number-nonconserving Bogoliubov approach.
In Ref.\cite{GIO} the fluctuations from collective excitations are
shown to follow the law $\left\langle \delta ^{2}N_{{\bf
{0}}}\right\rangle \sim N^{4/3}$. In this section, the Bogoliubov
theory will be developed based on our canonical statistics to
discuss the condensate fluctuations originating from collective
excitations. According to the Bogoliubov theory \cite{BOG,GIO2},
the total number of particles out of the condensate is given by

\begin{equation}
{N_{T}=\sum_{nl\neq 0}N_{nl}=\sum_{nl\neq 0}(u_{nl}^{2}+v_{nl}^{2})f_{nl}.}
\label{bthermal}
\end{equation}
\noindent The real quantities $u_{nl}$ and $v_{nl}$ satisfy the
relations

\begin{equation}
{u_{nl}^{2}+v_{nl}^{2}=\frac{\left[ \left( \varepsilon
_{nl}^{B}\right) ^{2}+g^{2}n_{0}^{2}\right] ^{1/2}}{2\varepsilon
_{nl}^{B}}},  \label{uv-1}
\end{equation}

\begin{equation}
{u_{nl}v_{nl}=-\frac{gn_{0}}{2\varepsilon _{nl}^{B}},}  \label{uv-2}
\end{equation}
\noindent where $f_{nl}$ is the number of the collective
excitations excited  in the system at the thermal equilibrium

\begin{equation}
{f_{nl}=\frac{1}{\exp \left[ \beta \varepsilon _{nl}^{B}\right] -1}}.
\label{bdistribution}
\end{equation}
\noindent In addition, the energy of the collective excitations
entering Eqs. (\ref{uv-1}) and (\ref{uv-2}) is given by the
dispersion law \cite{STR}

\begin{equation}
{\varepsilon _{nl}^{B}=\hbar \omega _{ho}\left( 2n^{2}+2nl+3n+l\right)
^{1/2}.}  \label{benergy}
\end{equation}
\noindent These phonon-like collective excitations are in
excellent agreement with the measurement of experiments. The
dispersion law (\ref
{benergy}) is valid if the conditions $N_{{\bf {0}}}a/a_{ho}>>1$ and $%
\varepsilon _{nl}<<\mu $ are satisfied. The contribution to the
condensate fluctuations due to these discrete low energy modes are
important
because $f_{nl},u_{nl}^{2}+v_{nl}^{2},u_{nl}v_{nl}\propto 1/\sqrt{%
2n^{2}+2nl+3n+l}$ at low excitation energies.

In Eq. (\ref{bthermal}) $N_{nl}$ can be regarded as the effective
occupation number of non-condensed atoms, while

\begin{equation}
{N_{nl}^{B}=\frac{N_{nl}}{u_{nl}^{2}+v_{nl}^{2}}=f_{nl}}  \label{bogo}
\end{equation}
\noindent is the occupation number of the collective excitations.
From the form of $f_{nl}$ one can construct the partition function
of the collective excitations in the frame of canonical ensemble

\begin{equation}
{Z_{B}=\sum_{\left\{nl\right\} }\exp \left[ -\beta \sum_{nl}
N_{nl}^{B}\varepsilon _{nl}^{B}\right].}  \label{bpar1}
\end{equation}
\noindent From Eq. (\ref{bogo}) $Z_{B}$ becomes

\begin{equation}
{Z_{B}=\sum_{\left\{ \Sigma N_{nl}=N\right\} } \exp \left[ -\beta
\sum_{nl}N_{nl}\varepsilon _{nl}^{eff}\right],}  \label{bpar2}
\end{equation}
\noindent where $\varepsilon _{nl}^{eff}= \varepsilon
_{nl}^{B}/\left( u_{nl}^{2}+v_{nl}^{2}\right) $ can be taken as an
effective energy level of the thermal atoms. In this case $Z_{B}$
is the partition function of a fictitious boson system, which is
composed of  $N$ non-interacting Bosons whose energy level is
determined by $\varepsilon _{nl}^{eff}$. From (\ref{bpar2}) the
most probable value $N_{{\bf {0}}}^{p}$ is given by

\begin{equation}
{N_{{\bf {0}}}^{p}=N-\sum_{nl\neq 0}\frac{1}{\exp \left[ \beta \left(
\varepsilon _{nl}^{eff}-\varepsilon _{nl=0}^{eff}\right) \right] -1}.}
\label{bmost}
\end{equation}
\noindent It is obvious that the occupation number of low $n,l$ in
Eq. (\ref{bmost}) coincides with that of Eq. (\ref{bthermal}).
Other $N_{{\bf {0}}}$ is determined by

\begin{equation}
{N_{{\bf {0}}}=N-\sum_{nl\neq 0}\frac{1}{\exp \left[ \beta \left(
\varepsilon _{nl}^{eff}-\varepsilon _{nl=0}^{eff}\right) \right] \exp \left[
-\alpha \left( N,N_{{\bf {0}}}\right) \right] -1}.}  \label{bother}
\end{equation}

From Eqs. (\ref{bmost}) and (\ref{bother}) we obtain

\begin{equation}
{\alpha \left( N,N_{{\bf {0}}}\right) \approx -}\frac{N_{{\bf {0}}}-N_{{\bf {%
0}}}^{p}}{\sum_{nl\neq 0}\left( u_{nl}^{2}+v_{nl}^{2}\right)
^{2}f_{nl}^{2}}. \label{balpha}
\end{equation}
\noindent When getting (\ref{balpha}) we have used the
approximation $f_{nl}\approx k_{B}T/\varepsilon _{nl}^{B}$ for low
energy collective excitations. Thus the probability distribution
function of the condensate is given by

\begin{equation}
{G_{B}\left( N,N_{{\bf {0}}}\right) =A}_{B}{\exp \left[ -\frac{\left( N_{%
{\bf {0}}}-N_{{\bf {0}}}^{p}\right) ^{2}}{2\sum_{nl\neq 0}\left(
u_{nl}^{2}+v_{nl}^{2}\right) ^{2}f_{nl}^{2}}\right] ,}  \label{bgauss-dis}
\end{equation}
where $A_{B}$ is a normalization constant. Therefore, the
condensate fluctuations due to the collective excitations reads

$${
\left\langle \delta^{2} N_{{\bf {0}}}\right\rangle_{collective}=
}$$

\begin{equation}
{   \frac{%
\sum_{N_{{\bf {0}}}=0}^{N}N_{{\bf {0}}}^{2}G_{B}\left( N,N_{{\bf {0}}%
}\right) }{\sum_{N_{{\bf {0}}}=0}^{N}G_{B}\left( N,N_{{\bf {0}}}\right) }-
\left[ \frac{\sum_{N_{{\bf {0}}}=0}^{N}N_{{\bf {0}}}G_{B}\left( N,N_{{\bf {0}%
}}\right) }{\sum_{N_{{\bf {0}}}=0}^{N}G_{B}\left( N,N_{{\bf {0}}}\right) }%
\right] ^{2}.}  \label{f-b}
\end{equation}
\noindent Eqs. (\ref{bgauss-dis}) and (\ref{f-b}) provide the
formulas for calculating  the condensate fluctuations originating
from the collective excitations.

Below the temperature $T_{m}$ which corresponds to the maximum
fluctuations, we obtain the analytical result for the condensate
fluctuations

$${
\left\langle \delta ^{2}N_{0}\right\rangle_{collective}=\frac{\pi ^{2}}{%
12\zeta \left( 2\right) }B\left( \frac{ma^{2}k_{B}T_{c}}{\hbar ^{2}}\right)
^{2/5}N^{4/3}=
}$$

\begin{equation}
{  \frac{1}{2}B\left( \frac{ma^{2}k_{B}T_{c}}{\hbar ^{2}}\right)
^{2/5}N^{4/3},}  \label{b-below}
\end{equation}

\noindent where $B$ is a dimensionless parameter, which is the
same as that obtained in Ref.\cite{GIO}. Note that compared with
the result obtained by Ref.\cite{GIO},  the coefficient in
(\ref{b-below}) differs by a factor $\frac{1}{2}$. The expression
(\ref{b-below}) shows clearly that the condensate fluctuations due
to the collective excitations are anomalous, {\it i.e.},
proportional to $N^{4/3}$. Note that $G_{B}\left( N,N_{{\bf
{0}}}\right) $ is a Gaussian distribution function, the anomalous
behavior of the condensate fluctuations comes from the factor
$2\sum_{nl\neq 0}\left(
u_{nl}^{2}+v_{nl}^{2}\right) ^{2}f_{nl}^{2}$, which is proportional to $%
N^{4/3}$.

At the critical temperature the probability distribution is given by $G_{B}\left( T=T_{c}\right) =\exp \left[ -N_{%
{\bf {0}}}^{2}/\gamma \right]$, where $\gamma =2\sum _{nl}\left(
u_{nl}^{2}+v_{nl}^{2}\right) ^{2}f_{nl}^{2}$. In this case, we obtains the
analytical result of the condensate fluctuations

\begin{equation}
{\left\langle \delta ^{2}N_{0}\right\rangle |_{T=T_{c}}=0.18\gamma=
0.18B\left( \frac{ma^{2}k_{B}T_{c}}{\hbar ^{2}}\right) ^{2/5}N^{4/3}.}
\label{bnear-trap}
\end{equation}
\noindent From (\ref{b-below}) and (\ref{bnear-trap}) we find that
the behavior of the condensate fluctuations based on the
Bogoliubov theory is rather different from that of the
lowest-order perturbation theory.


\section{Discussion and conclusion}


In this paper, a canonical ensemble approach has been developed to
investigate the mean ground state occupation number and condensate
fluctuations for interacting and non-interacting Bose gases.
Different from the conventional methods, the analytical
probability distribution function of the condensate has been
obtained directly from the partition function of the system. Based
on the probability distribution function, we have calculated the
thermodynamic properties of the Bose gas, such as the condensate
fraction and the fluctuations. Through the calculations of the
probability distribution function, we have provided a simple
method to recover the applicability of the saddle-point method for
studying the condensate fluctuations. In fact, the theory of the
improved saddle-point method  developed in this work can be
applied straightforwardly to consider the condensate fluctuations
in other physical systems, such as the interacting Bose gas
confined in a box \cite{xiong1}, the interacting Bose gas in
low-dimensions, etc.. The probability distribution function can
also be used to discuss other interesting problems, such as the
phase diffusion of the condensate.

For the harmonically trapped interacting Bose gas, we found that
different approximations for weakly interacting Bose gases give
quite different theoretical predictions concerning the condensate
fluctuations. In our opinion the lowest-order perturbation theory
gives in some sense the condensate fluctuations due to normal
thermal atoms, while the Bogoliubov theory gives the condensate
fluctuations originating from the collective excitations.  The
contributions to the condensate fluctuations due to the collective
excitations mainly come from the low energy modes, and it is
obvious that the condensate fluctuations based on the lowest-order
perturbation theory miss the contributions coming from the
collective excitations. Considering the fact that the
contributions due to low energy thermal atoms in the lowest order
perturbation theory is relatively small, the overall condensate
fluctuations may be written as

\begin{equation}
{\left\langle \delta ^{2}N_{{\bf {0}}}\right\rangle _{all}=\left\langle
\delta ^{2}N_{{\bf {0}}}\right\rangle _{int}+\left\langle \delta ^{2}N_{{\bf
{0}}}\right\rangle _{collective},}  \label{all-fluctuation}
\end{equation}
\noindent where $\left\langle \delta ^{2}N_{{\bf
{0}}}\right\rangle _{int}$ and $\left\langle \delta ^{2}N_{{\bf
{0}}}\right\rangle _{collective}$ are condensate fluctuations due
to the normal thermal atoms and the collective excitations,
respectively.


\section*{Acknowledgment}

This work was supported by the Science Foundation of Zhijiang
College, Zhejiang University of Technology and Natural Science
Foundation of Zhejiang Province. G. X. Huang was  supported by the
National Natural Science Foundation of China, the Trans-Century
Training Programme Foundation for the Talents and the University
Key Teacher Foundation of Chinese Ministry of Education. S. J. Liu
and H. W.  Xiong thank Professors G. S. Jia and  J. F. Shen for
their enormous encouragement.

\section*{Appendix}

In this appendix, the method of saddle-point integration described
by Darwin and Fowler \cite{DAR} is used to investigate the
partition function of the fictitious $N-N_{{\bf {0}}}$
non-interacting bosons. The partition function of the fictitious
system is given by

\begin{equation}
{Z_{0}\left( N_{T}\right) =\sum_{\sum_{{\bf {n}}\neq {\bf 0}}N_{{\bf {n}}%
}=N_{T}}\exp \left[ -\beta \sum_{{\bf {n}\neq 0}}N_{{\bf {n}}}\varepsilon _{%
{\bf {n}}}\right] ,}  \label{a-function-1}
\end{equation}

\noindent where $N_{T}=N-N_{{\bf {0}}}$ is the number of particles out of
the condensate.

Because of the restriction $\sum_{{\bf {n}\neq 0}}N_{{\bf {n}}}=N_{T}$ in
the summation of Eq. (\ref{a-function-1}), $Z_{0}\left( N_{T}\right) $ can
not be explicitly evaluated. To proceed we define a generating function for $%
Z_{0}\left( N_{T}\right) $ in the following manner. For any complex number $%
z $, we take

\begin{equation}
{G_{0}\left( T,z\right) =\sum_{N_{T}=0}^{\infty }z^{N_{T}}Z_{0}\left(
N_{T}\right).}  \label{a-generate-define}
\end{equation}

\noindent The generating function can be evaluated easily. The result of $%
G_{0}\left( T,z\right) $ is given by

\begin{equation}
{G_{0}\left( T,z\right) =\prod_{{\bf {n}}\neq {\bf 0}}\frac{1}{1-z\exp
\left[ -\beta \varepsilon _{{\bf {n}}}\right] }.}  \label{a-generate}
\end{equation}

To obtain $Z_{0}\left( N_{T}\right) $ we note that by definition $%
Z_{0}\left( N_{T}\right) $ is the coefficient of $z^{N_{T}}$ in
the expansion of $G_{0}\left( T,z\right) $ in powers of $z$.
Therefore we have

\begin{equation}
{Z_{0}\left( N_{T}\right) =\frac{1}{2\pi i}\oint dz\frac{G_{0}\left(
T,z\right) }{z^{N_{T}+1}},}  \label{a-relation}
\end{equation}

\noindent where the contour of integration is a closed path in the complex $%
z $ plane about $z=0$. Let $g\left( z\right) $ be defined by

\begin{equation}
{\exp \left[ g\left( z\right) \right] =\frac{G_{0}\left( T,z\right) }{%
z^{N_{T}+1}},}  \label{a-g-function}
\end{equation}

\noindent then $Z_{0}\left( N_{T}\right) $ becomes

\begin{equation}
{Z_{0}\left( N_{T}\right) =\frac{1}{2\pi i}\oint dz\exp \left[ g\left(
z\right) \right].}  \label{a-z-gfunction}
\end{equation}

The saddle point $z_{0}$ is determined by

\begin{equation}
{\frac{\partial g\left( z_{0}\right) }{\partial z_{0}}=0.}  \label{a-saddle}
\end{equation}

\noindent From Eq. (\ref{a-g-function}) we obtain

\begin{equation}
{N_{T}=z_{0}\frac{\partial }{\partial z_{0}}\ln G_{0}\left( T,z_{0}\right)
-1.}  \label{a-ttt}
\end{equation}

\noindent By Eq. (\ref{a-generate}), one gets

\begin{equation}
{N_{T}=\sum_{{\bf {n}}\neq {\bf 0}}\frac{1}{\exp \left[ \beta \varepsilon _{%
{\bf {n}}}\right] z_{0}^{-1}-1}.}  \label{a-thermal}
\end{equation}

\noindent Noting that Eq. (\ref{a-thermal}) is exactly the equation to
determine the number of condensed atoms within the grand canonical ensemble,
the saddle point $z_{0}$ can be also regarded as the fugacity of the
fictitious $N-N_{{\bf {0}}}$ non-interacting bosons.

Expanding the integrand of Eq. (\ref{a-z-gfunction}) about $z=z_{0}$, we have

$${
Z_{0}\left( N_{T}\right) =\frac{1}{2\pi i}\oint dz\exp \left[ g\left(
z_{0}\right) +\right.
}$$

\begin{equation}
{\left. \frac{1}{2}\left( z-z_{0}\right) ^{2}\frac{\partial ^{2}}{%
\partial z_{0}^{2}}g\left( z_{0}\right) +\cdots \right],}
\label{a-z-expansion}
\end{equation}

\noindent where

\begin{equation}
{\frac{\partial ^{2}}{\partial z_{0}^{2}}g\left( z_{0}\right) =\frac{%
G_{0}^{\prime \prime} \left( T,z_{0}\right) }{G_{0}\left( T,z_{0}\right) }-%
\frac{N_{T}^{2}-N_{T}}{z_{0}^{2}}.}  \label{a-g-second}
\end{equation}

\noindent By putting $z-z_{0}=iy$ we obtain

\begin{equation}
{Z_{0}\left( N_{T}\right) \approx \frac{\exp \left[ g\left( z_{0}\right)
\right] }{2\pi }\int_{-\infty }^{\infty }\exp \left[ -\frac{1}{2}\frac{%
\partial ^{2}}{\partial z_{0}^{2}}g\left( z_{0}\right) y^{2}\right] dy.}
\label{a-z-finial1}
\end{equation}

\noindent Thus we have

\begin{equation}
{Z_{0}\left( N_{T}\right) =\frac{G_{0}\left( T,z_{0}\right) }{%
z_{0}^{N_{T}+1} \left[ 2\pi g^{\prime \prime}\left( z_{0}\right) \right]
^{1/2}}.}  \label{a-z-finial}
\end{equation}
With these results the free energy $A_{0}\left( N,N_{{\bf
{0}}}\right) $ of the fictitious system is then given by

$${
A_{0}\left( N,N_{{\bf {0}}}\right) = -k_{B}T\left\{ \ln G_{0}\left(
T,z_{0}\right) -N_{T}\ln z_{0}\right.
}$$

\begin{equation}
{\left .-\ln z_{0}-\frac{1}{2}\ln \left[ 2\pi
g^{\prime \prime}\left( z_{0}\right) \right] \right\}.}  \label{a-free-1}
\end{equation}

\noindent In the case of $N_{T}>>1$, the last two terms in Eq. (\ref
{a-free-1}) can be omitted. Therefore

\begin{equation}
{A_{0}\left( N,N_{{\bf {0}}}\right) =-k_{B}T\left[ \ln G_{0}\left(
T,z_{0}\right) -N_{T}\ln z_{0}\right].}  \label{a-free-2}
\end{equation}

\noindent From Eq. (\ref{a-generate}), we obtain the relation between $%
A_{0}\left( N,N_{{\bf {0}}}\right) $ and $z_{0}$ of the fictitious
system:

\begin{equation}
{-\beta \frac{\partial }{\partial N_{{\bf {0}}}}A_{0}\left( N,N_{{\bf {0}}%
}\right) =\ln z_{0}.}  \label{a-free-relation}
\end{equation}

\noindent Eqs. (\ref{a-thermal}) and (\ref{a-free-relation}) are
useful relations used in the text.


\end{document}